# Igor TURKIN, Lina VOLOBUIEVA, Andriy CHUKHRAY, Oleksandr LIUBIMOV

*National Aerospace University "Kharkiv Aviation Institute", Kharkiv, Ukraine*

## MODELING CUBESAT STORAGE BATTERY DISCHARGE: EQUIVALENT CIRCUIT VERSUS MACHINE LEARNING APPROACHES

*The subject of the article is the study and comparison of two approaches to modelling the battery discharge of a CubeSat satellite: analytical using equivalent circuit and machine learning. The article aims to make a reasoned choice of the approach to modelling the battery discharge of a CubeSat satellite. Modelling the battery discharge of a satellite will enable the prediction of the consequences of disconnecting the autonomous power system and ensure the fault tolerance of equipment in orbit. Therefore, the selected study is relevant and promising. This study focuses on the analysis of CubeSat satellite data, based explicitly on orbital data samples of the power system, which include data available at the time of the article's publication. The dataset contains data on the voltage (mV), current (mA), and temperature (Celsius) of the battery and solar panels attached to the five sides of the satellite. In this context, two approaches are considered: analytical modelling based on physical laws and machine learning, which uses empirical data to create a predictive model. Results: A comparative analysis of the modeling results reveals that the equivalent circuit approach has the advantage of transparency, as it identifies possible parameters that facilitate understanding of the relationships. However, the model is less flexible to environmental changes or non-standard satellite behaviour. The machine learning model demonstrated more accurate results, as it can account for complex dependencies and adapt to actual conditions, even when they deviate from theoretical assumptions. However, the model requires prior training on a large amount of data and is less well understood in terms of physical laws. General conclusions. The equivalent circuit approach provides high accuracy and reliability under known conditions, but it is limited when external parameters change. The machine learning approach demonstrates better overall accuracy and stability, especially under variable or unpredictable conditions, but requires a large amount of high-quality data and more complex interpretation. Thus, the most effective approach may be a hybrid one, where the analytical model serves as the basis and machine learning is used as a tool for refining or compensating for inaccuracies.*

## 1. Introduction

Due to its attractive cost, the availability of commercially ready-made solutions, and a relatively short implementation time, CubeSat has gained popularity among space researchers. It is currently used to solve a wide range of tasks. The general overview for the current state-of-the-art SmallSat technologies [1] state the growing popularity of small satellites in general and CubeSats in particular, and show that since 2013, the flight heritage for small spacecraft has dramatically increased and has become the main primary source of access to space for commercial, government, private, and academic.

The CubeSat project was initiated in 1999 by scientists from California Polytechnic State University and Stanford University's Space Systems Development Laboratory. Specification [2] defines a 1U (U stands for 'Unit') CubeSat as a small satellite that has a standard size and form, which is a 10 cm cube with a mass of up to 2 kg. A CubeSat can consist of several units. The current version of the specification describes the design of CubeSats up to 12U.

### 1.1. Motivation

According to various estimates, the global CubeSat market will show a GAGR over 15% in the coming years (according to CubeSat Market Research Report https://straitsresearch.com/report/cubesat-market Straits research expects it to reach USD 1,305.56 million by 2032, with GAGR of 15.1% during the forecast period (2024-2032) with base year 2023 while by IMARC Group in its report "CubeSat Market Size, Share, Trends and Forecast by Size, Application, End User, Subsystem, and Region", 2025-2033" expects the market to reach USD 1,608.98 Million by 2033, exhibiting a CAGR of 16.3% during 2025-2033 with base year 2024.

According to [5], the most significant number of CubeSat missions were with a mass of 3U (~45.5% of the total and 53.3% of the successfully launched). CubeSats are being assigned more and more complex tasks, which increases the requirements for their capabilities. Fig. 1 shows the dynamics of the deployment of missions based on CubeSat over the last 20 years. In the previous ten years, missions based on 12U (53 missions, first in 2016), 16U (26 missions, first in 2019), and the first 20U mission in 2023 (China, Tianzhi-2D) have been successfully launched. Although the development of CubeSat projects is fast and accessible to a wide range of researchers, providing opportunities for implementing both commercial and educational projects [3], the failure rate of such projects is relatively high [4]. Fig. 2 shows the success (i.e., how successful it was, not the current state) of such satellites for educational and scientific missions from 2003 to 2025 (May). To construct the visual image, data from the nanosatellite and CubeSat database [5] were utilized; the database's last significant update was on April 30, 2025. The paper presents a comparative analysis of analytical and machine learning-based approaches to modeling CubeSat storage battery discharge.

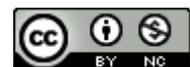

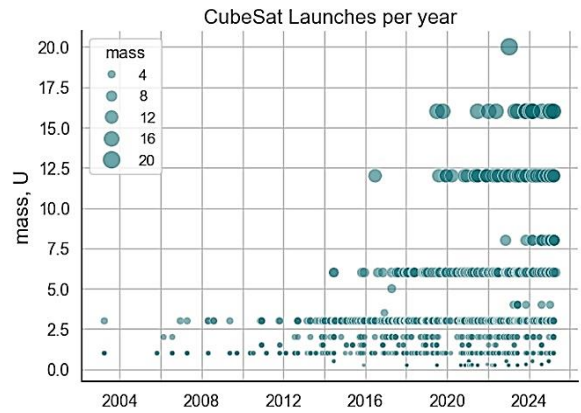

Fig. 1. Dynamics of the launches based on CubeSat missions for the period from 2003 to 2025

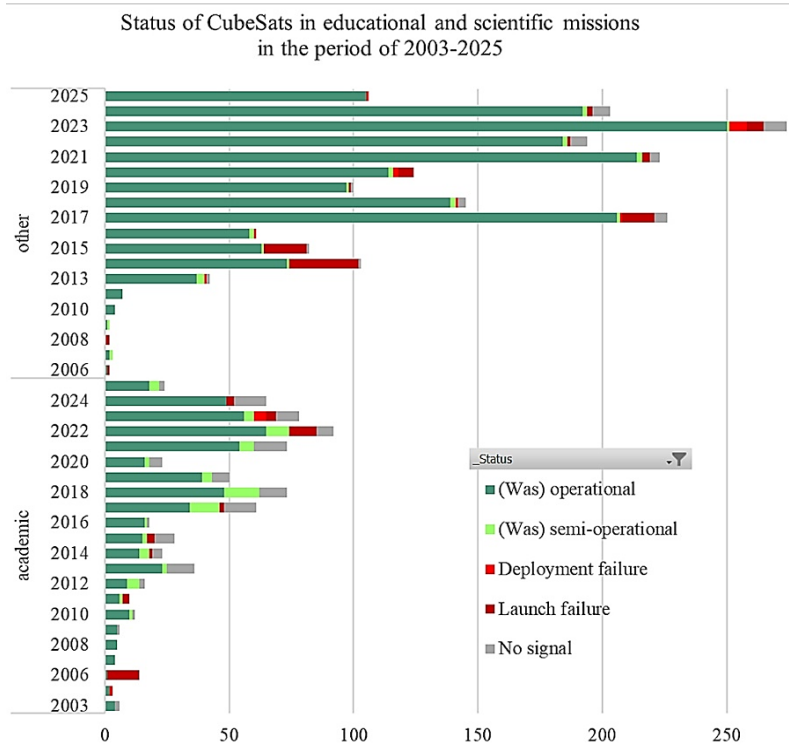

Fig. 2. Status of CubeSats in educational and scientific missions in the period of 2003 to 2025 (May)

## 1.2. State of the art

According to UNOOSA (United Nations Office for Outer Space Affairs) records, 13,469 satellites are orbiting Earth as of April 2025, of which only 12,205 satellites are active (as of late April 2025), as the satellite tracking website "Orbiting Now", that maintains the records of the satellites in various Earth orbits.

The report [6] analyzed the reasons for the partial and complete failure of missions with small satellites. According to the report, between 2000 and 2016, 41.3% of mission launches with small satellites failed partially or entirely. Of these, 24.2% of missions suffered complete failure, 11% failed partially, and launch vehicle failures accounted for another 6.1%. In work [7], based on the analysis of failure reports from 2000 to 2012, among the most frequent causes of mission failures are configuration or interface issues between communication hardware (27%), the Electrical Power System (EPS, 14%), and the flight processor (6%). Therefore, it is crucial to predict satellite failures through modelling.

Several papers have investigated ways to improve energy harvesting using solar panels and techniques such as maximum power point tracking (MPPT), with a particular focus on the benefits of implementing machine learning techniques to enhance the performance of Electrical Power Systems (EPS). Thus, in paper [8], a deep learning-based MPPT approach is proposed to improve CubeSat power generation; the authors used annual data obtained from simulations of a 3U CubeSat to train the model. The authors of the study [9] provide an overview of various types of MPPT methods, including classical, intelligent, optimization-based, and hybrid methods.

Authors consider EPS, and we didn't find other research connected with CubeSat storage battery discharge modeling, except [10]. However, there is no comparative analysis of analytical and machine learning approaches to modeling.

The authors of the study [11] investigated the statistical reliability of small satellites using empirical failure data for the period 2010-2020 and showed a general trend that reflects that at the beginning of the mission the probability of failure of each subsystem is high and constantly decreases during the first two years, then the values gradually decrease and fluctuate around the nominal value. In particular, in the work [11] it is stated that the contribution of the failure of the power subsystem to the satellite failure after a certain specific time in orbit is determined as follows: after 30 days – 17.63%, after one year – 11.78%, after two years – 7.42%, after 10 years – 9.97%.

Study [12] (as of October 2024) states that of the 2,714 CubeSats launched, 677 have experienced problems or failures, not all of which could have resulted in mission loss. According to [12], most failures were caused by launch and deployment failures. Among the problems identified, communication failures, power system failures, and high spin rates were noted as the most common.

## 2. Objectives and approach

In this study, the primary focus is on analyzing the Electrical Power System (EPS) of 1U CubeSat satellites based on on-orbit data samples from the TSURU satellite dataset [13], which includes data from its deployment into orbit to the present time.

This work investigates the possibility of using machine learning to predict the battery discharge of a CubeSat satellite. By the aim of the study, the following tasks must be solved:

1. Analysis of available observational data and handling outliers (section 3).

2. CubeSat storage battery discharge modeling using equivalent circuit techniques. Evaluation of the model accuracy. (section 4.1).

3. CubeSat storage battery discharge modeling based on machine learning approach (section 4.2). It contains data preprocessing (described in section 4.2.1) and further machine learning model training (section 4.2.2). Evaluation of the built machine learning model performance (section 4.2.3).

4. Comparative analysis of the obtained results by the two approaches used (section 5).

The source data is based on the BIRDS open-source standardized bus [13]. The dataset contains voltage (mV), current (mA), and temperature (in degrees Celsius) data for the storage battery and solar panels attached to the five sides of the satellite. This data is collected by the onboard computer every 90 seconds in normal mode or every 10 seconds in fast sampling mode. The dataset contains data on solar panels and batteries from the time when they were launched into orbit until the end of life of the UGUISU, RAAVANA, and NEPALISAT satellites. The TSURU satellite dataset contains data since its launch into orbit and will continue to be collected throughout its lifetime.

The sampling interval for UGUISU, RAAVANA, and NEPALISAT was 5 seconds, and for TSURU, it was 10 seconds. NEPALISAT, RAAVANA, and UGUISU operated in orbit for more than two years before their re-entry. TSURU was still operating in orbit when the dataset was released.

Thus, if the solar panel generates too much voltage for the battery, it is limited by the DC/DC converter to 4.2V, preventing overcharging. The generated energy is stored in a battery pack consisting of six rechargeable Eneloop Nickel Metal Hydride (NiMH) batteries, each with a minimum capacity of 1900 mAh, arranged in a 3-series and 2-parallel configuration. A thermistor is installed between the batteries to measure the temperature. Since low temperatures negatively affect battery capacity, the battery is wrapped with Kapton tape and a polyimide heater to help maintain thermal balance. The Electrical diagram is shown in Fig. 3. The dataset is missing the solar cell current, $I_{sra}$, and the load current, $I_{raw}$.

The usual practice of using a storage battery in a satellite is that when there are signs of approaching full discharge of the storage battery (SB), it is necessary to urgently turn off everything superfluous to preserve the viability of the whole satellite. Excess primarily includes the payload; only the life support systems remain switched on. This certainly reduces the beneficial effect of using the satellite if the load limiting mode is turned on early. Still, such a mode should be activated automatically, as waiting for a human operator to make a decision would result in the loss of the entire satellite.

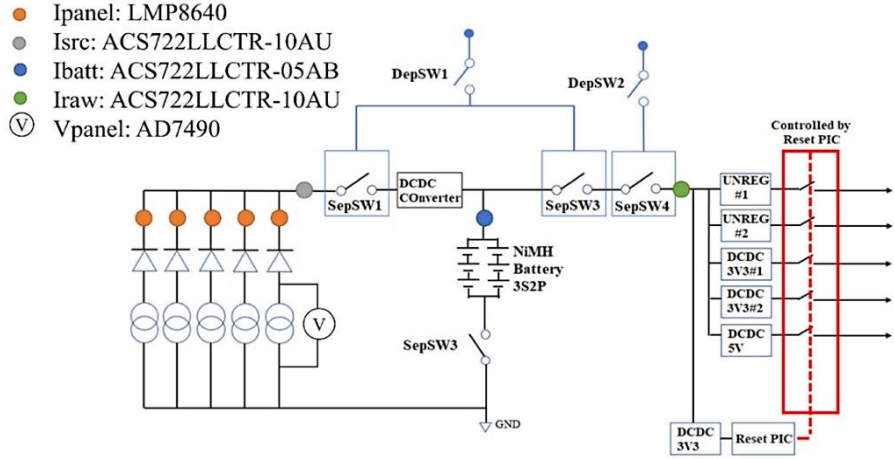

Fig. 3. BIRDS-4 satellite EPS block diagram [13].

Characteristics of the photoconverter are shown in Table 1. The charging and discharge characteristics of the manufacturer of each battery are shown in Fig. 4. A sign of the exhaustion of the battery's capacity is the end of the linear section of the discharge characteristic of the SB. Still, this moment of termination depends significantly not only on the current used to discharge the chemical battery, but also on the history of its operation. It is known that control charge-discharge cycles, which can be carried out in shadowless areas of the orbit, allow not only the estimation of the current capacity of the SB but also the restoration of its characteristics.

Table 1
Dependence of photoconverter characteristics
on equivalent radiation dose,
source: AZUR SPACE Solar Power GmbH

|  | BOL | 2.5E14 | 5E14 | 1E15 |
|---|---|---|---|---|
| Average Open Circuit $V_{oc}$ [mV] | 2690 | 2606 | 2554 | 2512 |
| Average Short Circuit $I_{sc}$ [mA] | 519.6 | 517.9 | 513.4 | 501.3 |
| Voltage at max. Power $V_{mp}$ [mV] | 2409 | 2343 | 2288 | 2244 |
| Current at max. Power $I_{mp}$ [mA] | 502.9 | 501.7 | 499.1 | 485.1 |
| Average Efficiency $\eta_{bare}$ (1367 W/m²) [%] | 29.3 | 28.4 | 27.6 | 26.3 |
| Average Efficiency $\eta_{bare}$ (1353 W/m²) [%] | 29.6 | 28.7 | 27.9 | 26.6 |
| Standard: CASOLBA 2005 (05-20MV1, etc); Spectrum: AMO WRC = 1367 W/m²; T = 28 °C | | | | |

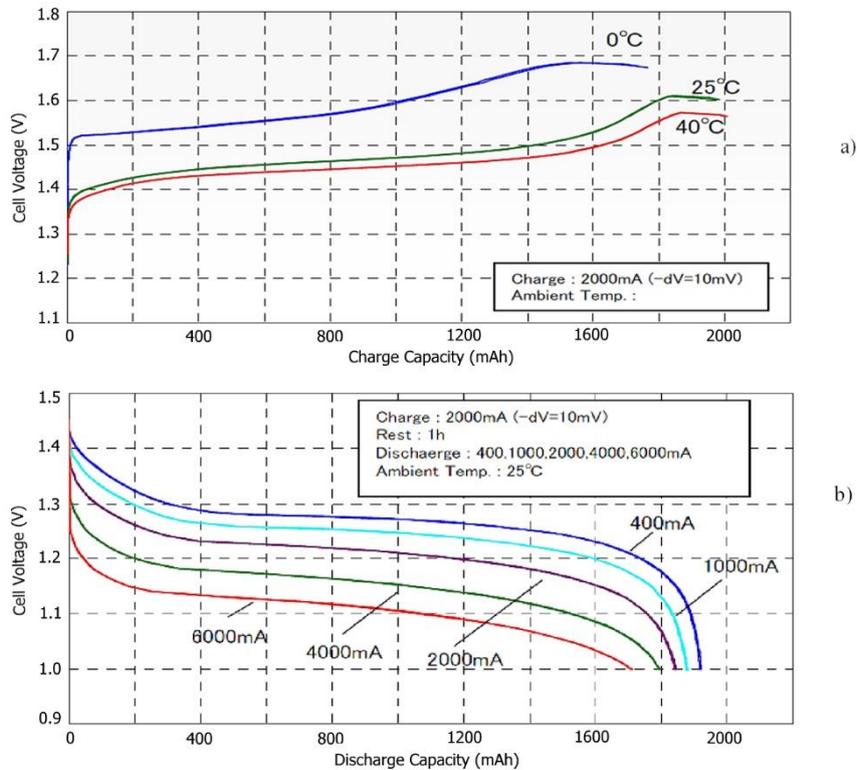

Fig. 4. Charge (a) and Discharge (b) characteristics of each battery, measured in laboratory conditions,
supplied by the manufacturer

According to the dataset, the average current of the SB discharge is approximately 300 mA, but the current can briefly increase to a maximum value of 1200 mA.

The BIRDS satellites were launched from the International Space Station (ISS) into the ISS orbit (altitude 400 km, inclination angle: 51.6 °, duration of one revolution: 92.6 min). In the orbit of the ISS, the Solar Beta Angle (the height of the Sun above the plane of the orbit) varies within ±75.1 ° during the year (Fig. 5a, c), affecting the percentage of time during which the satellites are illuminated, which determines the generation of electricity by solar panels and the temperature profile.

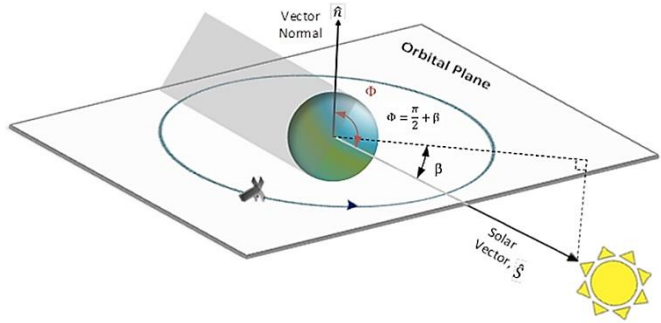

a) Solar Beta Angle definition

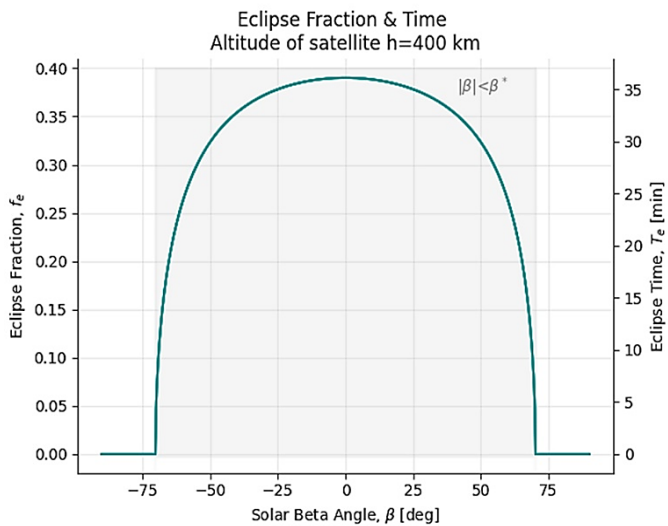

b) Dependency of the Eclipse Fraction and Eclipse Time on the Solar Beta Angle

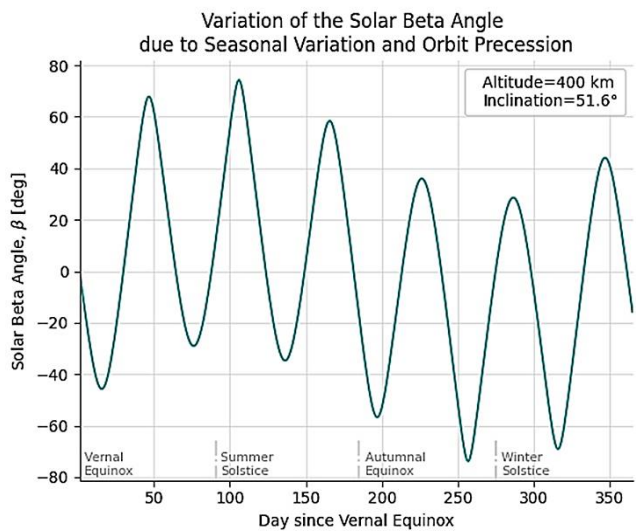

c) Solar Beta Angle variation over a year (possible profile)

Fig. 5. The periods during which the Satellites are illuminated due to the ISS orbit

When the absolute value of the Solar Beta Angle exceeds 69.9 °, the satellite is in a shadowless orbit (Fig. 5 b) [14]. The satellites of this series have no orientation control, so the satellites rotate freely at a speed of about 3 deg/s on each axis. Five of the six faces of the cubic-shaped satellite provide a power supply.

The orientation and illumination of the solar panels are determined by the satellite's geometry, as shown in Fig. 6a. When the satellite is on the illuminated part of the orbit, the best energy supply will be if it is deployed at 45 ° along two axes. The difference is about 2-2.2 times (Fig. 6 b).

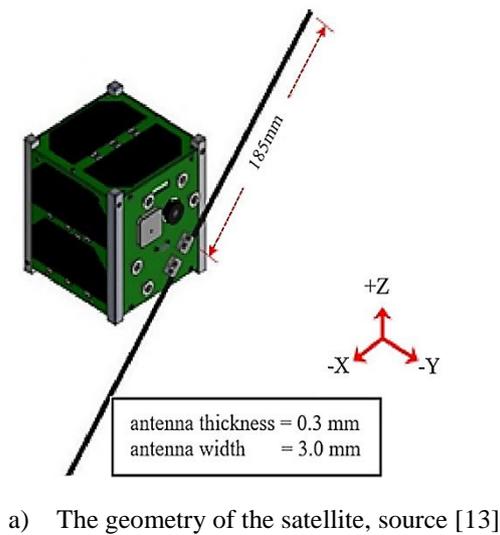

a) The geometry of the satellite, source [13]

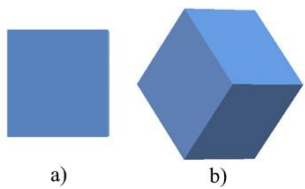

b) Satellite orientation for the worst (a) and the best (b) power harvesting

Fig. 6. Appearance of the satellite

When the device leaves the shadow area, the cooled PB generates maximum energy, which decreases by 50–80% within 5–7 minutes and then changes according to the temperature variation of the photoconverters.

## 3. Data analysis

The available data [13] covers the period from March 2021 to January 2022 (namely for dates: March 28, April 20, April 27, May 13, May 21, May 28, June 01, June 25, June 11, June 17, July 05, July 09, August 07, August 22, September 07, September 03, September 11, September 13, September 19, September 25, October 03, October 15, November 02, November 16, November 05, November 22, November 27, December 07, December 20, December 25 and January 24, 2022) for the following characteristics:

— Time stamp – the time at which the data sample was measured (sec);

— temperatures of the five surfaces of the photovoltaic cells (°C);

— output voltages of the five photovoltaic cells (mV);

— generated current from the five photovoltaic cells (mA);

— voltage (V), charge-discharge current (mA), and temperature (°C) of the energy storage device.

The dataset is valuable and important because it includes in-orbit data collected by four different CubeSat satellites that share the same bus system structure: the NEPALISAT, RAAVANA, UGUISU, and TSURU satellites. This data can be used to estimate the energy available in orbit for a 1U CubeSat, assessing the feasibility of missions. The Battery discharge periods were extracted from this dataset. Data visualization for 20_02_2021 (complete set a) and selected discharge period b)) are shown in Fig. 7.

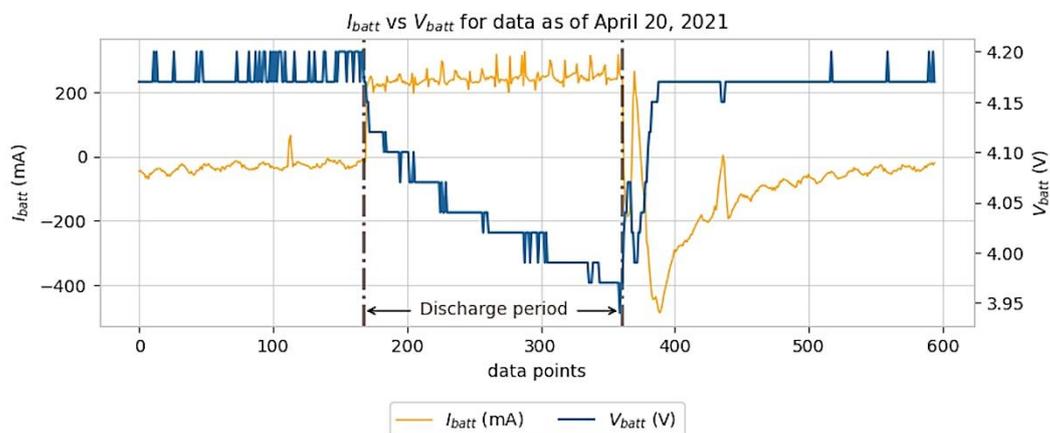

Fig. 7 Data visualization of the whole set and the April 20, 2021, discharge period.

For a better understanding of their composition and range of changes, statistical information was collected, and the DOD was calculated (described in detail on p. 4.1) and presented in Table 2.

Table 2
Data set features description (5709 measures)

|  | V<sub>batt</sub> (V) | I<sub>batt</sub> (mA) | DOD (mA*hour) | T<sub>batt</sub> (°C) |
|---|---|---|---|---|
| mean | 4.0298 | 263.0029 | 65.9986 | 6.2552 |
| std | 0.0575 | 76.0496 | 42.8586 | 2.7049 |
| min | 3.79 | 5.64 | 0 | 1.07 |
| max | 4.2 | 1200.42 | 225.3272 | 15.63 |

During data analysis, a significant deviation in the $I_{batt}$ current values was detected in the data sets for the dates 01/24/2022, 07/05/2021, 07/09/2021, and 05/21/2021, where $I_{batt}$ reached values above 1000 mA (Fig. 8). Also, several data points that represents negative $I_{batt}$ i.e., battery charge instants in the middle of the discharge periods, were found in the discharge data set. Fig. 8 shows the distribution of both typical and untypical (i.e. "too low" – negative $I_{batt}$ values and "too high" – $I_{batt}$ > 500 mA) $I_{batt}$ data.

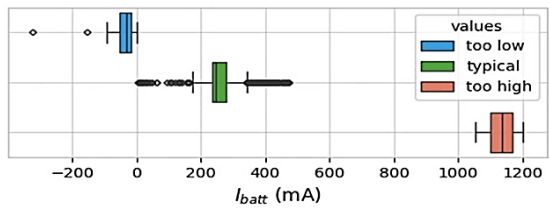

Fig. 8. $I_{batt}$ values distribution visualization for the whole discharge data set.

# 4. CubeSat storage battery discharge modeling

In this study, CubeSat battery discharge modeling is performed using two approaches: equivalent circuit and machine learning, and the obtained results are analyzed.

### 4.1. Equivalent circuit based model of the CubeSat storage battery discharge

To build the model, the efficiency of energy transfer from solar batteries to a chemical battery was first determined according to the criterion of the minimum variance of dependency.

$$N_{Load}(\tau) = N_{SB}(\tau) + \eta \sum_{1}^{5} V_{p_i}(\tau) \cdot I_{p_i}(\tau) \qquad (1)$$

Energy conversion efficiency η =92.6 %. To confirm the correctness of the calculations, the graph shown in Fig. 9.

The satellite design information has been released as a standardized, open-source BIRDS bus system for the fast and easy development of satellites for educational and research purposes. The data can be used as a standard data set to verify the on-orbit performance of a satellite power system developed using the BIRDS bus.

Usually, the discharge characteristics of a storage battery are of fundamental importance when calculating the energy supply system of satellites (Fig 10). The manufacturer indicates in its own specifications.

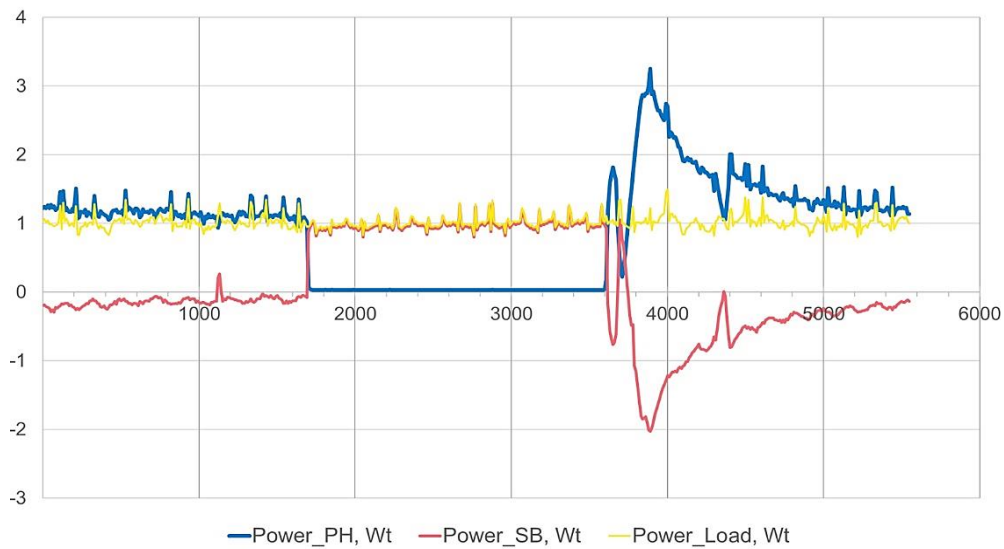

Fig.9. Confirmation the correctness of the calculations

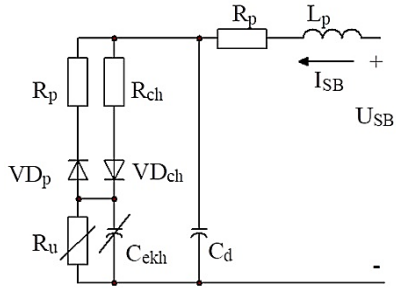

Fig.10. The equivalent circuit of the storage battery

In Figure 10:
– Rp, Lp – resistance and inductance of external connections, electrodes, and electrolyte;
– Rp, VDp formalize the nonlinear polarization effect as a function of the charge level, discharge current, and battery temperature;
– Rch, VDch are similar to Rp, VDp for the charge mode;
– Cd is the capacitance of the double layer;
– Cekh is the battery capacity as an electrochemical storage; -
– Ru – a non-linear resistor that simulates the processes associated with self-discharge and the occurrence of side chemical reactions.

A simplified mathematical model of discharge characteristics contains the dependence of the depth of discharge as an integral on the discharge current and the battery voltage, which depends on the current, temperature, and depth of discharge:

$$DOD(\tau) = \int_0^\tau I_d(t)\,dt$$

$$U(\tau) = U_0 - K_{DOD}(T) \cdot DOD(\tau) - R(T) \cdot I_d(\tau) - \quad (2)$$

$$U_p(T) \cdot \exp\left(-\frac{DOD(\tau)}{DOD_{kp}(T)}\right)$$

Identification of the parameters of the discharge characteristic model is the determination of the structure and parameters of the temperature:

$K_{DOD}(T)$ – the angle of inclination of the linear section of the discharge characteristic;

$R(T)$ – internal resistance;

$U_p(T)$ – change in voltage on the initial non-linear section of the discharge characteristic;

$DOD_{kp}(T)$ – the duration of the initial nonlinear section of the discharge characteristic.

Obtained results based on the considered approach. To analyze the obtained results, the accuracy of the model was calculated using the complete sets; the results are presented in the graphs (Figs. 11).

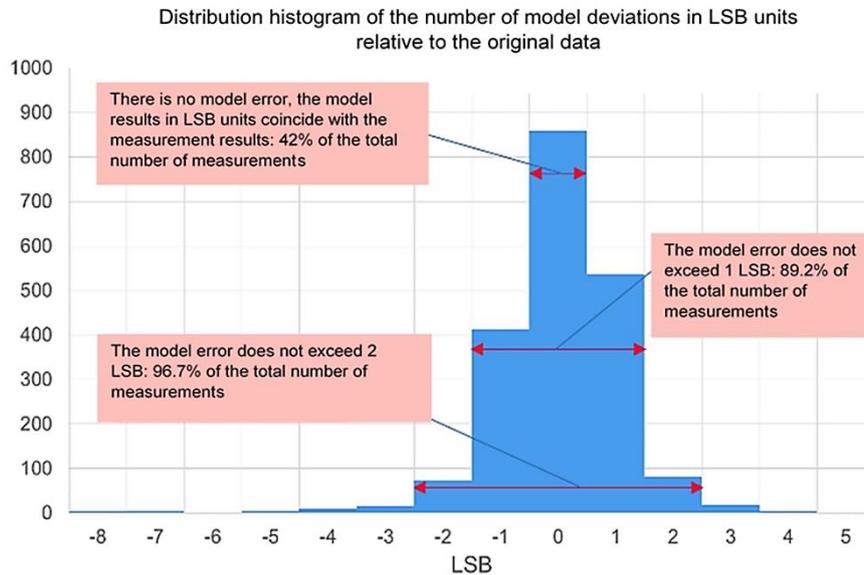

Fig. 11. Accuracy of the model on the complete data set
(31-bit characteristics)

## 4.2. Machine learning based model of the CubeSat storage battery discharge

Based on the results of the data analysis described in Section 3, a dataset was generated for further training of the model, containing data corresponding to battery discharge periods. The training data set comprises data on battery discharge periods collected from all available sets (March 2021 to January 2022) for both typical and atypical $I_{batt}$ values. The data set features description for the whole set and for the typical $I_{batt}$ values is represented in Table 2 and Table. 3.

Table 3
Data set features description for the typical $I_{batt}$ values (5024 measures)

|       | $V_{batt}$ (V) | $I_{batt}$ (mA) | DOD (mA*hour) | $T_{batt}$ (°C) |
|-------|----------------|-----------------|---------------|-----------------|
| mean  | 4.0305         | 258.6566        | 64.9712       | 6.1894          |
| std   | 0.0547         | 42.7227         | 40.8895       | 2.7484          |
| min   | 3.87           | 5.64            | 0             | 1.07            |
| max   | 4.2            | 474.18          | 181.4189      | 15.63           |

**Handling Outliers.** Data outliers (data points where a short-term increase of $I_{batt}$ was observed in the middle of the discharge period) were excluded from the set.

**Feature Extraction.** Taking into account the correlation of the accessible data, the following features were selected: $I_{batt}$ – the current value of the battery in calibrated format (mA), and $T_{batt}$ – the temperature value of the battery in calibrated format (°C), and extracted from the full data set for periods of satellite battery discharge. The data series for all selected features has no missing values.

### 4.2.1. Feature engineering

**Feature Creation.** The feature DOD (discharge characteristic, mA*hour) was created using the equation from (2). The selected features correlation study was performed on the whole set for the battery discharge period, the results are shown in Fig. 12.

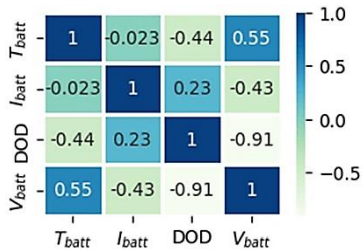

Fig. 12. Visualization of the correlation of dataset features.

As can be seen from the figure, the input data are correlated with the target feature and are not overcorrelated. New feature matrix consisting of all polynomial combinations of the both selected and created features with degree less than or equal to the 8 was generated.

**Scaling.** Each selected feature was scaled and transformed individually according to the equation (3).

$$X_{scaled} = \frac{X - X_{min}}{X_{max} - X_{min}} \quad (3)$$

The target feature (label) is $V_{batt}$ – Voltage value of the battery in calibrated format (V).

### 4.2.2. Machine learning model creation and training

For CubeSat storage battery discharge prediction, the cross-validated Lasso regressor model (LassoLarsCV), which is based on the least-angle regression (LARS) algorithm [15] and can overcome multicollinearity, was chosen. A cross-validation estimator was selected due to its ability to support warm-starting by reusing precomputed results from previous steps of the cross-validation process, as well as to provide the advantage of the best training/development data set split. The metric used for the model performance evaluation – R2 Score, defined as

$$R^2 = 1 - \frac{\sum_{i=1}^{N}\left(y_i - \hat{y}_i\right)^2}{\sum_{i=1}^{N}\left(y_i - \bar{y}\right)^2}, \quad \bar{y} = \frac{1}{N}\sum_{i=1}^{N} y_i \quad (4)$$

where: $\hat{y}_i$ is the predicted value of $V_{batt}$ for the i-th sample and $y_i$ - the corresponding actual value of $V_{batt}$.

To chain multiple estimators into one, a pipeline (Python, Scikit-learn) was developed. The pipeline (Fig. 14) consists of the following steps:

1) data normalization using MinMaxScaler with default feature range [0,1];

2) generation a new feature matrix consisting of all polynomial combinations of the features with degree less than or equal to 8-degree using PolynomialFeatures;

3) regressor LassoLarsCV, cv=6.

Before using cross-validation to ensure unbiased model performance estimation, the dataset was split into training and test sets at a ratio of 80:20 %.

A characteristic feature of LARS is its computational efficiency; the algorithm requires the same order of calculations as the ordinary least squares (OLS) method [16]. The LassoLarsCV finds the relevant regularization parameter (alpha) values itself, which can help prevent the model from overfitting. The model's performance (cross-validation R² score) on the training set is 0.924, and on the test set, it is also 0.924.

### 4.2.3. Model results analysis

To analyze the model's performance on the entire dataset, ensuring that the model generalizes appropriately and does not exhibit signs of overfitting, a graph is used. (Fig. 13, 14) of the model's predicted values on the full range of values for $I_{batt}$ at T=5°C and DOD in the range of 0-200 mA*hour with a step of 20 were constructed.

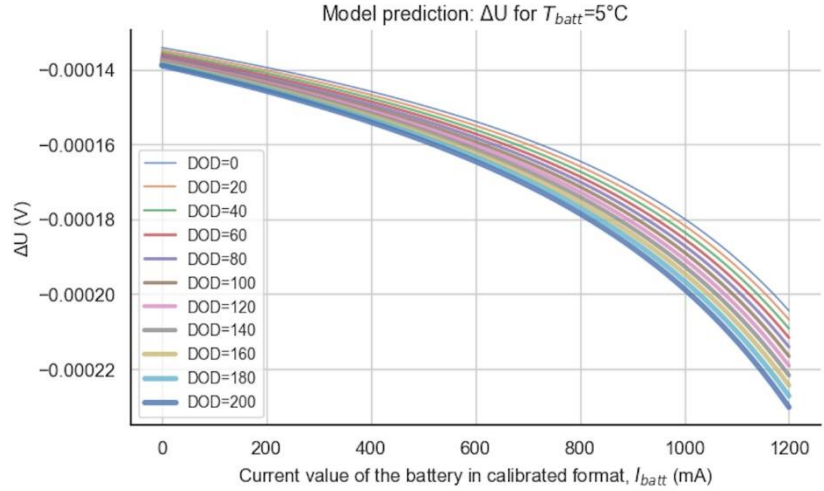

Fig. 13. Dependency graph for $\frac{\Delta U}{\Delta I}$

Shown in the visual, ΔU was calculated according to equation (5):

$$\frac{\Delta U}{\Delta I} = \frac{f\left(I + \Delta I, DOD, T\right) - f\left(I, DOD, T\right)}{\Delta I} \quad (5)$$

U for the visual was calculated using equation (6) for the whole range of values for $I_{batt}$, i.e., $\overline{\left(0,1200\right)}$ mA:

$$U = f\left(I, DOD, T\right) \quad (6)$$

where the temperature setpoint T = 5℃ and DOD varies as $DOD = \left\{\overline{\left(0, 200\right)} \, mA * hour, \, step = 20\right\}$.

The constant temperature test was chosen to accurately analyze the model's predictions for battery behaviour under different orbital conditions.

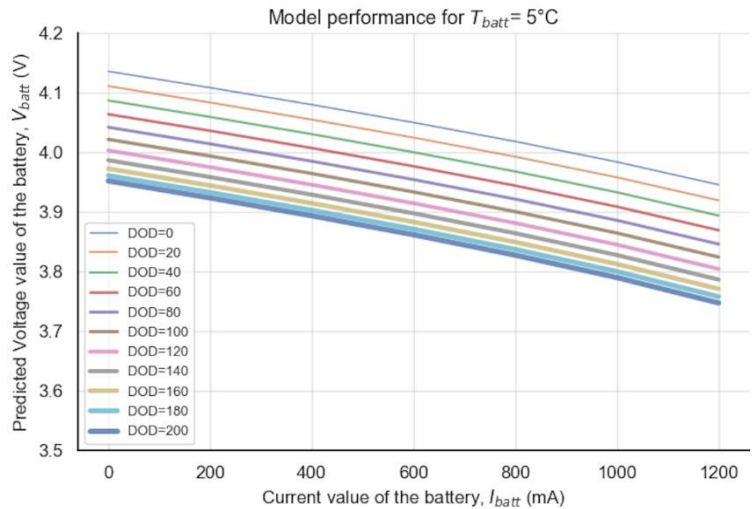

Fig. 14. Dependency graph for $V_{batt}$

To analyze the model prediction error, histograms were constructed as shown in Fig. 15.

The histograms show the error of the model in LSB. The data predicted by the model with an accuracy of the fifth sign were used to construct the histograms.

The maximum absolute error is 2.7488 LSB. The percentage of the model prediction error above 2 LSB is 0.44%.

## 5. Comparative analysis of the results of analytical modeling and modeling based on machine learning

The accuracy of both models was assessed based on the analysis of histograms of the distribution of deviations of model calculations relative to experimental data in the most minor bit units – LSB (Least Significant Bit). This approach enables us to quantitatively evaluate the accuracy of models and assess their suitability for practical applications.

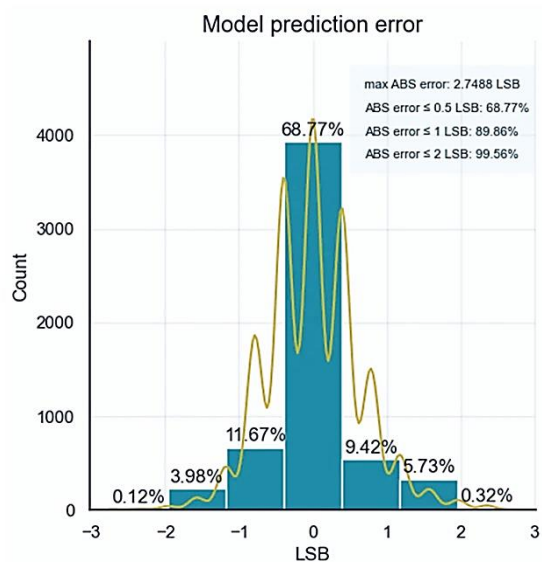

Fig. 15. Model prediction error histograms

As for the 31-bit analytical model, one of the most important indicators of its quality is that in 42% of all modeling cases, the model result entirely coincides with real measurements, meaning the error is zero. This indicates that in almost half of all cases, the model entirely accurately reflects the behavior of the object. Such a result is especially valuable in space research, where accuracy plays a key role in ensuring the smooth operation of the equipment.

Another important aspect is the level of permissible deviations. In 89.2% of cases, the model error does not exceed 1 LSB, which demonstrates a high approximation to actual data. This indicates that the model is reliable in most situations and can be used in scenarios where minimal deviation is acceptable.

The most convincing argument in favor of the model is that in 96.7% of all measurements the error does not exceed 2 LSB. This means that only in 3.3% of cases can the model give a deviation that, although insignificant, still exceeds this threshold. This indicator indicates the exceptional stability of the model and its high efficiency in predicting the actual discharge process.

Thus, the histogram analysis allows us to conclude that the constructed model is very accurate and efficient. It is capable of reflecting complex physical processes with a high degree of reliability and can be successfully utilized in satellite systems where accuracy and reliability are crucial.

The same conclusions can be drawn for the 27-bit analytical model.

As for the machine learning (ML) model, it is a modern approach that allows taking into account complex nonlinear dependencies and ensuring high prediction accuracy. Analysis of the results, based on the histogram of the distribution of absolute deviations in the least significant bits (LSB) relative to the experimental data, allows us to assess the quality of the constructed ML model.

First of all, the maximum absolute error (Max abs error) is only 2.7488 LSB, which is already a sign of a very moderate level of mistakes. In the conditions of satellite technologies, where every unit of measurement is essential, such an error margin is acceptable and technically justified.

The most impressive is the distribution of model accuracy:

• 68.77% of all forecasts have an absolute error that does not exceed 0.5 LSB. This means that almost 70% of the model results are practically identical to exact measurements. For energy monitoring tasks, this demonstrates the model's extraordinary sensitivity and accuracy.

• 89.86% of the predictions have an error of no more than 1 LSB, which confirms the stability and predictability of the model for most discharge scenarios.

• Finally, 99.56% of all results have an error of no more than 2 LSB, which indicates almost complete agreement with real data - deviations exceeding this threshold are extremely rare and can be associated with anomalous or extreme conditions.

Thus, the results show that the machine learning model demonstrates high accuracy with a minimal level of error. Such a model is a reliable tool for predicting the state of the satellite battery, capable of adapting to actual data and maintaining accuracy even under conditions of complex dynamic changes.

## 6. Conclusions and Discussions

Thus, mathematical modeling of the satellite battery discharge is a critically important task for ensuring the stable operation of the spacecraft. In this context, two approaches are considered: analytical modeling, which is based on physical laws, and machine learning, which utilizes empirical data to create a predictive model.

The equivalent circuit approach has the advantage of transparency, as it allows for clear tracing of the relationships between parameters; however, the model is less flexible in responding to changes in the environment or non-standard behavior of the system.

The machine learning model demonstrated even better results. This indicates higher accuracy and the model's ability to account for complex dependencies and adapt to real-world conditions, even when they deviate from theoretical assumptions. However, the model requires prior training on a large amount of data and is less interpretable in terms of physical laws.

General conclusions:

• The equivalent circuit approach provides high accuracy and is reliable under known conditions, but is limited when changing external parameters.

• The machine learning approach demonstrates better overall accuracy and stability, especially under variable or unpredictable conditions, but requires large amounts of qualitative data and more complex interpretation.

Thus, the most effective approach in practical applications may be a hybrid one, where the analytical model serves as the basis and machine learning (ML) is used as a tool for refining or compensating for inaccuracies.

**Contributions of authors:** conceptualization, formal analysis – **Igor Turkin**; methodology – **Igor Turkin, Lina Volobuieva, Oleksandr Liubimov**; software – **Igor Turkin, Lina Volobuieva**; original draft preparation – **Igor Turkin, Lina Volobuieva**; review and editing – **Igor Turkin, Lina Volobuieva, Oleksandr Liubimov**; supervision – **Igor Turkin**; comparative analysis – **Andriy Chukhray**; abstract, conclusions – **Andriy Chukhray**; design of the article – **Andriy Chukhray**.

## Funding

This research is carried out in the frame and on a budget of the national Ukrainian grant project NRFU.2023.04/0143 - "Experimental development and validation of the on-board computer of a dual-purpose unmanned aerial vehicle."

## Conflicts of Interest

The authors declare no conflict of interest.

All authors have read and agreed to the published version of the manuscript.

**ПОРІВНЯЛЬНИЙ АНАЛІЗ ПІДХОДІВ НА ОСНОВІ ІНТЕРВАЛЬНОЇ МАТЕМАТИКИ І МАШИННОГО НАВЧАННЯ ДО МОДЕЛЮВАННЯ РОЗРЯДУ БАТАРЕЇ СУПУТНИКА CUBESAT**

*І.Б. Туркін, Л.О. Волобуєва, А.Г. Чухрай, А.В. Любімов*

Предметом статті є дослідження та порівняння двох підходів до моделювання розряду акумулятора супутника CubeSat: аналітичного з використанням еквівалентної схеми та машинного навчання. Метою статті є обґрунтований вибір підходу до моделювання розряду акумулятора супутника CubeSat. Моделювання розряду акумулятора супутника дозволить передбачити наслідки відключення автономної енергосистеми та забезпечити відмовостійкість обладнання на орбіті. Тому обране дослідження є актуальним та перспективним. Це дослідження зосереджено на аналізі даних супутника CubeSat, що базується на орбітальних зразках даних енергосистеми, які включають наявні дані на момент опублікування статті. Набір даних містить дані про напругу (мВ), струм (мА) та температуру (за Цельсієм) акумулятора та сонячних панелей, прикріплених до п'яти боків супутника. У цьому контексті розглянуто два підходи: аналітичне моделювання на основі фізичних законів та машинне навчання, яке використовує емпіричні дані для створення прогнозної моделі. Результати. Порівняльний аналіз результатів моделювання показує, що моделювання за допомогою методу еквівалентних схем має перевагу прозорості, оскільки такий підхід визначає можливі параметри, що сприяють розумінню взаємозв'язків. Однак модель менш гнучка до змін навколишнього середовища або нестандартної поведінки супутника. Модель машинного навчання продемонструвала точніші результати, оскільки вона може враховувати складні залежності та адаптуватися до фактичних умов, навіть коли вони відхиляються від теоретичних припущень. Однак модель вимагає попереднього навчання на великій кількості даних і менше зрозуміла з точки зору фізичних законів. Загальні висновки. Моделювання за допомогою методу еквівалентних схем забезпечує високу точність і надійність за відомих умов, але такий підхід обмежений при зміні зовнішніх параметрів. заснований на використанні машинного навчання, демонструє кращу загальну точність і стабільність, особливо за змінних або непередбачуваних умов, але вимагає великої кількості високоякісних даних та більш складної інтерпретації. Таким чином, найефективнішим підходом може бути гібридний, де аналітична модель служить основою, а машинне навчання використовується як інструмент для уточнення або компенсації неточностей.

**Ключові слова:** малий супутник, CubeSat; Електроенергетична система; моделювання; машинне навчання.

**Туркін Ігор Борисович** – д-р техн. наук, проф., зав. каф. інженерії програмного забезпечення, Національний аерокосмічний університет «Харківський авіаційний інститут», Харків, Україна.

**Волобуєва Ліна Олексіївна** – канд. техн. наук, доцент., доцент каф. інженерії програмного забезпечення, Національний аерокосмічний університет «Харківський авіаційний інститут», Харків, Україна.

**Чухрай Андрій Григорович** – д-р техн. наук, проф., проф. каф. інженерії програмного забезпечення, Національний аерокосмічний університет «Харківський авіаційний інститут», Харків, Україна.

**Любімов Олександр Вікторович** – аспірант каф. інженерії програмного забезпечення, Національний аерокосмічний університет «Харківський авіаційний інститут», Харків, Україна.

**Ihor Turkin** – Doctor of Technical Sciences, Professor, Head of the Department of Software Engineering, National Aerospace University "Kharkiv aviation institute", Kharkiv, Ukraine,
e-mail: i.turkin@khai.edu, ORCID: 0000-0002-3986-4186.

**Lina Volobuieva** – Candidate of Technical Sciences, Associate Professor, Associate Professor at the Department of Software Engineering, National Aerospace University "Kharkiv Aviation Institute", Kharkiv, Ukraine,
e-mail: l.volobuieva@khai.edu, ORCID: 0000-0002-3466-5743.

**Andriy Chukhray** – Doctor of Technical Sciences, Professor, Professor of the Department of Software Engineering, National Aerospace University "Kharkiv aviation institute", Kharkiv, Ukraine,
e-mail: a.chukhray@khai.edu, ORCID: 0000-0002-8075-3664.

**Oleksandr Liubimov** - Ph.D. Student of the Department of Software Engineering, National Aerospace University "Kharkiv aviation institute", Kharkiv, Ukraine,
e-mail: oleksandr.liubimov@gmail.com, ORCID: 0000-0003-3636-6939.